# Machine Learning based Laser Failure Mode Detection

Khouloud Abdelli[1,2], Danish Rafique[1], and Stephan Pachnicke[2]

[1] ADVA Optical Networking SE, Fraunhoferstr. 9a, 82152 Munich/Martinsried, Germany
[2] Christian-Albrechts-Universität zu Kiel, Kaiserstr. 2, 24143 Kiel, Germany
E-mail: KAbdelli@advaoptical.com

**ABSTRACT**
Laser degradation analysis is a crucial process for the enhancement of laser reliability. Here, we propose a data-driven fault detection approach based on Long Short-Term Memory (LSTM) recurrent neural networks to detect the different laser degradation modes based on synthetic historical failure data. In comparison to typical threshold-based systems, attaining 24.41% classification accuracy, the LSTM-based model achieves 95.52% accuracy, and also outperforms classical machine learning (ML) models namely Random Forest (RF), K-Nearest Neighbours (KNN) and Logistic Regression (LR).
**Keywords**: Degradation, laser, reliability, fault detection, machine learning, recurrent neural networks

## 1. INTRODUCTION

Traditionally lasers have been widely used in various industrial sectors such as astronomy, communications, metrology and medical examinations. Since their inception, lasers have evolved in terms of optical output power and wavelength tuning range, among other aspects, allowing for various applications imposing stringent engineering requirements. Consequently, quantification of laser reliability has become a much more challenging issue [1]. To address this challenge, different reliability modelling and prediction approaches have been developed, ranging from empirical methods, based on the statistical analysis of the historical failure data, like Telcordia SR-322[2] to physical failure models. In contrast to the standard-based prediction methods the physical models allow the inclusion of the explicit impact of design, manufacturing, and operation on reliability [3]. While the accuracy has been enhanced, challenges like the costs and the complexity of failure modelling exist, together with the need for detailed manufacturing information and a knowledgeable team of experts.

In order to deal with the shortcomings of the different reliability models, data-driven fault-detection approaches based on ML techniques have been recently proposed. Trained on normal and faulty data, these models can detect and predict the failure rate, and further classify the type of the anomaly. In this respect, Rafique et al [4, 5] proposed a cognitive assurance fault detection architecture for optical networks. On the laser reliability aspect, Okaro et al [6] presented a fault detection approach for selective laser melting using semi-supervised ML.

In this paper, we detect and predict different laser failure modes using ML models with fast execution speed and high classification accuracy of up to 95.52%. While machine learning has been consistently used for industrial maintenance use cases targeting advanced predictive analytics, the application of machine learning methods for laser failure mode detection has not been explored yet. Fed with historical failure data, various machine learning algorithms such as Random Forest (RF), Logistic Regression (LR), Long Short-Term Memory (LSTM) and K-Nearest Neighbours (KNN) are trained and evaluated on test data containing partial failure data. The results show that ML-based models outperform the standard threshold-based system.

This contribution aims to (1) compare the efficiency of several machine learning techniques for detecting laser failure modes, and (2) to prove that the ML models outperform the conventional methods such as threshold-based fault detection approaches.

The paper is organized as follows. The synthetic data generation and the review of the various machine learning algorithms used for classification are described in Section 2. The comparison of various machine learning algorithms as well as the comparison of the chosen ML model with the threshold-based system are given in Section 3. The paper is concluded in Section 4.

## 2. SETUP & CONFIGURATIONS

### 2.1 Synthetic Data Generation

We generated synthetic data for our study targeting different laser failure modes, namely gradual, rapid and sudden degradation (Fig. 1) as well as normal laser operation. Equation (1) shows the variation of the operational current as a function of the time at constant power [7].

$$I(t) = I_0 + I_{nr}(t) \ where \ I_{nr}(t) = \beta \exp(k.t) \ and \ k = P^n \exp(\mu_0 - \frac{E_A}{k_B T}) \quad (1)$$

The parameter $n$ denotes the de-rating exponent, $E_A$ the activation energy, $\mu_0$ the scale parameter, $\beta$ the non-radiative current and $T$ the temperature. The input features, including optical power $P$, the threshold current $I_0$ and temperature $T$, are extracted from real laser datasheets specifications [8], whereas the underlying coefficients to create the different degradation patterns are generated using normal distributions shown in Fig. 2.

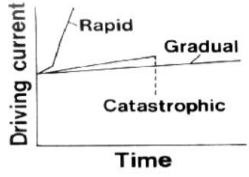 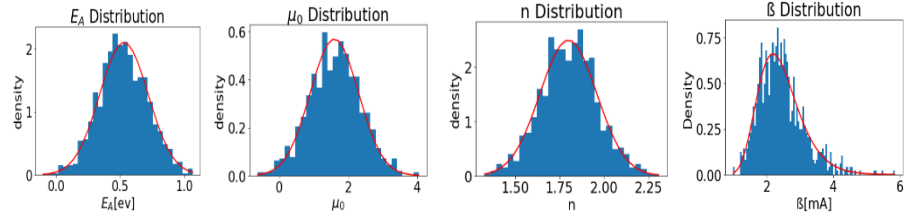

*Figure 1. Laser degradation modes [9]*  *Figure 2. Parameter distributions*

For each type of laser of degradation mode, as well as for normal laser behavior, approximately 1,500 samples are generated. In total, a dataset composed of 6,000 univariate time series was built to develop the laser fault detection model based on the variation of the current as a function of the time. The investigated laser parameters included optical power, current threshold, wavelength and temperature and the type of laser degradation.

### 2.2    Preprocessing

The dataset modeled the laser degradation modes at three different time scales: rapid degradation is observed within the first 100 hours of operation, gradual degradation could extend to several hundreds of hours, and catastrophic degradation appears after many hours of normal operation [10]. We normalized the said data to a window size of 100. The gradual degradation time series were compressed by averaging and in order to build a compact 100 sequence representative of this failure mode.

Figure 3 shows the different laser failure modes patterns after preprocessing, where the *x*-axis represents different time indices.

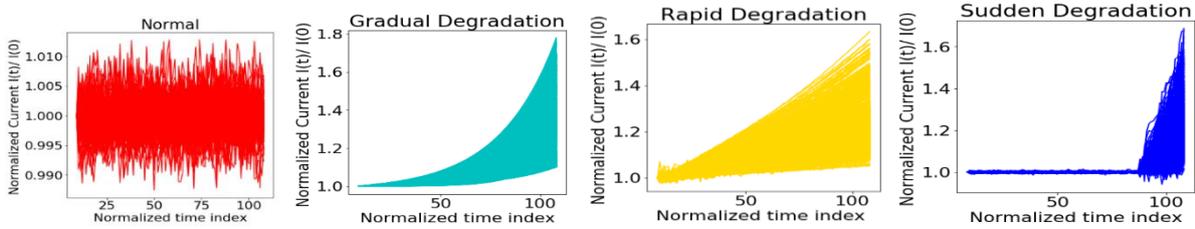

*Figure 3. Laser failure mode patterns*

After preprocessing, the data was divided into a 60% training dataset, a 20% validation dataset and a 20% test dataset. To mimic the real in-field data, the test dataset was modified in such a way that for each failure mode sample only 20 – 40% of the fault pattern is kept and prepended with a normal laser operation sequence representing 60% – 80% of the observation.

### 2.3    ML Algorithms

We represent laser failure mode detection as a univariate time series classification problem and use LSTM (long Short-Term Memory) network as the algorithm of choice, suitable to model the time information.

#### 2.3.1    LSTM-based fault detection model

LSTM, proposed by Hochreiter & Schmid Huber in 1997 as a solution to the gradient vanishing problem [11], is a special type of Recurrent Neural Network (RNN). The core computational unit of LSTM is called block memory, compromised of weights and three gates, governing the flow information to the cell state, namely the input, the forget and the output gate, noted as *i*, *f* and *o,* respectively. The gates are sigmoid functions, whereas *U* is the weights' matrix, *W* is the recurrent connection at the previous hidden layer and the current layer, and $h_{t-1}$ is the previous hidden layer. *C* is a "candidate" hidden state computed based on the current input and the previous hidden states. $C_{t-1}$ is the internal memory of the unit, a combination of the previous memory, multiplied by the forget gate, and the newly computed hidden state, multiplied by the input gate. Figure 4 shows the structure of an LSTM cell as well as the equations describing the functionality of LSTM gates.

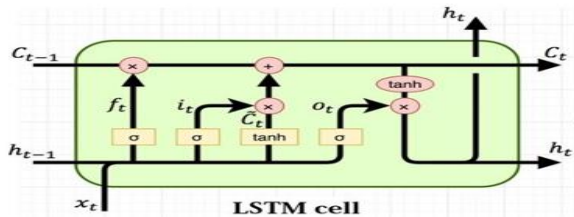

$$f_t = \sigma(\ x_t\, U^f\ +\ h_{t-1}\, W^f\ )$$
$$i_t = \sigma(\ x_t\, U^i\ +\ h_{t-1}\, W^i\ )$$
$$\tilde{C}_t = \tanh(\ x_t\, U^g\ +\ h_{t-1}\, W^g\ )$$
$$C_t = \sigma(\ f_t * C_{t-1}\ +\ i_t * \tilde{C}_t\ )$$
$$o_t = \sigma(\ x_t\, U^o\ +\ h_{t-1}\, W^o\ )$$
$$h_t = o_t * \tanh(\ C_t\ )$$

*Figure 4. LSTM structure and relevant relationships [12]*

In the training phase, current samples combined with laser parameters influencing the degradation like the current threshold, temperature, optical power and the wavelength were fed at the input nodes, and the type of the laser degradation (0: normal, 1: gradual degradation, 2: rapid degradation and 3: sudden degradation) was fed to the output node of LSTM . A categorical cross-entropy function was used as the loss function to update the weights of the LSTM model based on the error between the predicted and the desired output. Furthermore, the hyperparameter tuning of the model led to the selection of the RMSProp optimizer among Adam and Adadelta optimizers. The hidden state dimensionality was set to be 100 units, and the number of hidden layers was 2.

### 2.3.2    Baseline ML algorithms

The performance of the developed LSTM model was also compared with the classical ML classification algorithms, RF, K-NN and Multinomial LR, based on the same dataset used to train the LSTM model. To enhance the performance of these algorithms, a selection of the best hyperparameters was done through the grid search method. For KNN, the parameter $K$ was chosen to be 6. For LR, the regularization parameter $C$ was selected to be 100, and finally for RF, the number of trees was set to be 100.

## 3.    RESULTS AND DISCUSSION

To evaluate the performance of the different ML models four metrics were used: the accuracy, the recall, the precision and the F1 score. Figure 5 shows the detailed results of the comparison.

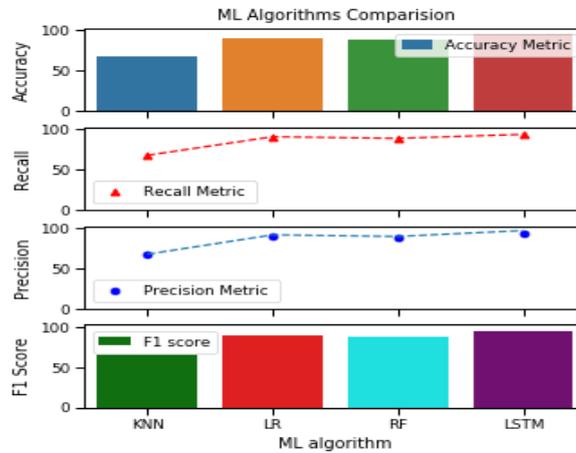

*Figure 5. Performance comparison of the different ML models*

The results of the experiment showed that the LSTM-based model was the best model in terms of all the evaluation metrics as LSTM is able to learn long term dependency in a temporal pattern. Multi class LR and RF performed as second and third best models, whereas KNN had the worst accuracy (Fig. 5).

The summary of the selected LSTM model performance is shown in Figure 6. The confusion matrix shows that misclassification rates. Receiver Operating Characteristic (ROC) curves summarize the trade-off between the true positive and false positive rates for the LSTM model using different probability thresholds whereas precision-recall curves summarize the trade-off between the true positive rate and the positive predictive value.

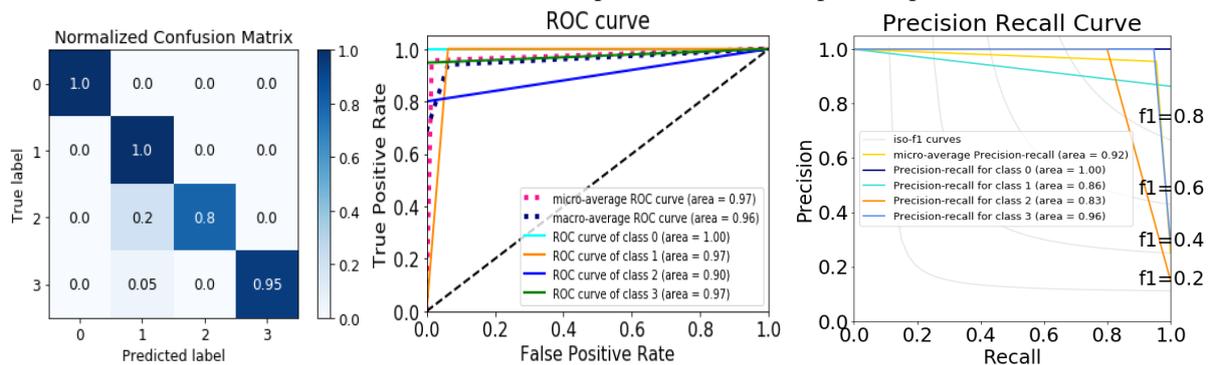

*Figure 6.  LSTM performance evaluation summary*

In order to compare the performance of the chosen LSTM-based fault detection model with the conventional laser failure detection methods, a multi-threshold system assigned different thresholds based on failures, EOL (end of life) criteria and laser specification and design. Tested on data containing partial test failure cases, the LSTM model outperformed the threshold system in terms of two metrics: (1) fault accuracy interpreted as the accurate detection of the failure, (2) and the classification accuracy defined as the accurate identification of the type of the laser degradation (Figure 7). The LSTM model was able to detect the laser failure with 99% fault accuracy and to recognize the class of the laser degradation mode with 95.52% classification accuracy. On the other hand, the threshold-based system's fault and classification accuracy were only 24.41%.

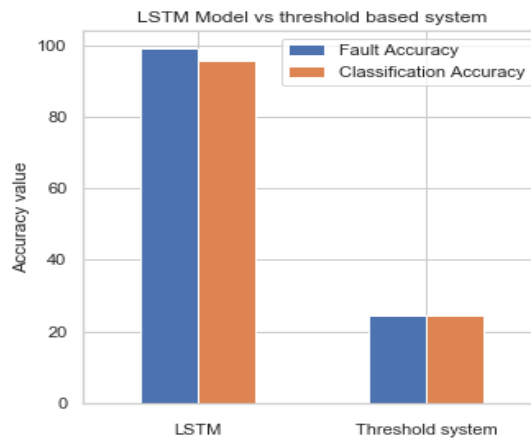

*Figure 7. LSTM Model vs Threshold based system comparison*

## 4. CONCLUSIONS

In this paper we proposed ML-based detection of laser failure mode based on LSTM recurrent neural networks. The LSTM-laser failure detection outperformed three different state-of-the-art classification ML models RF, KNN and LR, as well as the multi-threshold-based system in terms of classification accuracy. Future work will include the collection of real (experimental) or in-field data for the performance evaluation of the developed model.

## ACKNOWLEDGEMENTS

This work was funded from METRO-HAUL (G.A. nº 761727) and SENDATE SECURE-DCI (C2015/3-4).

## REFERENCES

[1] P. W. Epperlein: Semiconductor Laser Engineering, Reliability and Diagnostics, New York: John Wiley & Sons Ltd, 2013.

[2] Telcordia: Reliability Prediction for Electronic Equipment, Special Report SR-332, Issue 2, 2006.

[3] M. White, et al.: Microelectronics Reliability, JPL Publication 08-5 2/08, California, 2008.

[4] D. Rafique, et al.: Cognitive Assurance Architecture for Optical Network Fault Management, IEEE J. Lightw. Technol. **36**, 1443-1450, 2018.

[5] D. Rafique, et al.: Machine Learning for Network Automation, IEEE JOCN **10**, D126-D143, 2018.

[6] Okaro, et al.: Automatic Fault Detection for Laser Powder-Bed Fusion using Semi-Supervised Machine Learning, Additive Manufacturing, 2019.

[7] Haussler, et al.: Degradation Model Analysis of Laser Diode, JOMS: Materials in Electronics, 2008.

[8] Laser Lab Source :LASER DIODE SOURCE , available at https://www.laserdiodesource.com/laser-diodes-filtered-by-wavelength/1550nm-laser-diodes .

[9] J. G. McInerney: Experimental Description of Semiconductor Lasers , 2001, available at http://www.physics.ucc.ie/pres/SS01-L1-statics/sld073.htm .

[10] J. Jimenez: Laser diode reliability: Crystal defects and degradation modes , Comptes Rendus Physique - C R PHYS. 4. 663-673, 2003.

[11] Hochreiter, et al.: Long Short-Term Memory, Neural Computation, vol. 9(8), 1735–1780, 1997.

[12] S. Varsamopoulos, et al.: Decoding small surface codes with feedforward neural networks, Quantum Science and Technology, vol. 3, no. 1, p. 015004, 2018.